\documentstyle[preprint,tighten,aps,psfig]{revtex}
\def\ZETA{z}
\newcommand{\real}{{\sf I}\kern-.12em{\sf R}}

\begin{document}
\draft
\preprint{IFUP-TH 23/98, UCY-PHY 7/98}
\title{
Resummation of Cactus Diagrams in Lattice QCD}
\author{H. Panagopoulos$^a$, E. Vicari$^b$}
\address{
$^a$Department of Natural Sciences, University of Cyprus,
P.O.Box 537, Nicosia CY-1678, Cyprus. haris{@}europe.ns.ucy.ac.cy}
\address{
$^b$Dipartimento di Fisica dell'Universit\`a 
and I.N.F.N., 
Piazza Torricelli 2, I-56126 Pisa, Italy. vicari{@}mailbox.difi.unipi.it}


\maketitle

\begin{abstract}
We show how to perform a resummation, to all orders in perturbation theory,
of a certain class of gauge invariant diagrams in Lattice QCD. These diagrams
are often largely responsible for lattice artifacts. Our resummation leads to 
an improved perturbative expansion. Applied to a number of cases of interest,
this expansion yields results remarkably close to corresponding 
nonperturbative estimates. 


\end{abstract}

\newpage


\section{Introduction}
\label{introduction}

Ever since the earliest days of lattice field theory, one problem
present in most numerical simulations has been the calculation of
corrections induced by renormalization on Monte Carlo
results. Although this notorious problem has not as yet been
adequately dealt with, several methods have been used to address it:
To begin with, perturbation theory provides in principle a methodical
means of calculating, order by order in the coupling, renormalization
functions, operator mixing coefficients, etc. Its drawbacks lie in its
asymptotic nature, and in that it is a formidable task on the lattice,
which places severe limitations on the order to which it can be
carried out; indeed, at present, exact calculations in perturbative
lattice QCD reach only two loops (for 2-point
diagrams)~\cite{lw234,afp,cfpv}  and three loops (for vacuum
diagrams)~\cite{afp2}. Various 
nonperturbative, numerical approaches to renormalization functions
have also been devised and there has been recent progress both in
their range of applicability and in their
precision~\cite{romani,M-P-S-T-V,luescher}. Finally, much effort has
also gone in studying 
improved actions (which may, among other advantages, show improved
renormalization behaviour)~\cite{clover,lueschercl} and improved/boosted
perturbation theory~\cite{lepage}.

\noindent
\begin{minipage}{0.68\linewidth}
\ \ \ \ \ \ In this paper, we present an 
improvement of lattice perturbation
theory, which results from a resummation to all orders of a certain
class of diagrams, dubbed ``cactus'' diagrams. Briefly stated, these
are tadpole diagrams which become disconnected if any one of their
vertices is removed (see Figure 1). Our original motivation was the well known
observation of ``tadpole dominance'' in lattice perturbation theory
(see, e.g.~\cite{dashen}). 
This observation must clearly be taken with
a grain of salt: 
One-sided inclusion of tadpoles can ruin desirable
partial cancellations between tadpole and non-tadpole diagrams; worse,
\end{minipage}\hskip0.02\textwidth
\begin{minipage}{0.30\linewidth}
\medskip
\hskip1.0cm\psfig{figure=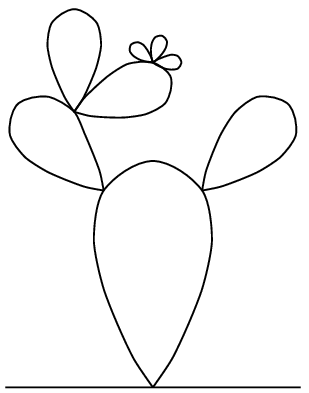,width=3truecm}\hskip1.0cm
\bigskip
{\centerline{\bf Figure 1: A cactus}}
\end{minipage}
their contribution is gauge dependent. 
The class of terms we propose
to resum circumvents the latter objection since, as we shall see, it
is gauge invariant; it also overcomes the former objection in known
cases. 

Cactus resummation may be applied either to bare quantities or to
quantities which have been calculated to a given order in perturbation
theory; thus contributions which are not included in the resummation
can be reintroduced in a systematic manner.

In Section~\ref{sec2} we present our calculation, leading to expressions for a
dressed propagator and dressed vertices of interest; some derivations
and technical details are relegated to the Appendices. In Section~\ref{sec3}, we
proceed to use these expressions to calculate various renormalization
functions, and compare our results with other methods: We find a
remarkable improvement in many cases. 

\section{The calculation}
\label{sec2}

\subsection{The dressed propagator}
\label{sec2.1}

Consider the standard Wilson action for $SU(N)$ lattice gauge fields:
\begin{equation}
S = {1\over g_0^2} \sum_{x,\mu\nu} {\rm Re}\, {\rm tr} \left(1-U_{x,\mu\nu}^\Box
\right)
\end{equation}
$U_{x,\mu\nu}^\Box$ is the usual product of link variables around a
plaquette in the $\mu{-}\nu$ plane with origin at $x\,$; in standard
notation it reads:
\begin{equation}
U_{x,\mu\nu}^\Box = e^{i g_0 A_{x,\mu}} e^{i g_0 A_{x+\mu,\nu}} e^{-i
g_0 A_{x+\nu,\mu}} e^{-i g_0 A_{x,\nu}} , \qquad A_{x,\mu} =
A_{x,\mu}^a T^a
\end{equation}
By the Baker-Campbell-Hausdorff (BCH) formula we have:
\begin{eqnarray}
U_{x,\mu\nu}^\Box &=& \exp\left\{i g_0 (A_{x,\mu} + A_{x+\mu,\nu} -
A_{x+\nu,\mu} - A_{x,\nu}) + {\cal O}(g_0^2) \right\} \nonumber \\
&\equiv& \exp\left\{ i g_0 F_{x,\mu\nu}^{(1)} +  i g_0^2
F_{x,\mu\nu}^{(2)} +  i g_0^3 F_{x,\mu\nu}^{(3)} + {\cal O}(g_0^4)
\right\}
\end{eqnarray}

The diagrams which we propose to resum to all orders will be cactus
diagrams made of vertices containing $F_{x,\mu\nu}^{(1)}\,$. Let us
see how such diagrams will dress the gluon propagator; we write:
\begin{equation}
\psfig{figure=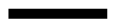,width=1truecm}
= \psfig{figure=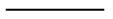,width=1truecm} +
\psfig{figure=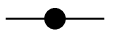,width=1truecm} +
\psfig{figure=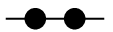,width=1truecm} + \cdots
\label{propdress}
\end{equation}
where the one-particle irreducible piece is given by the recursive
equation:
\begin{equation}
\psfig{figure=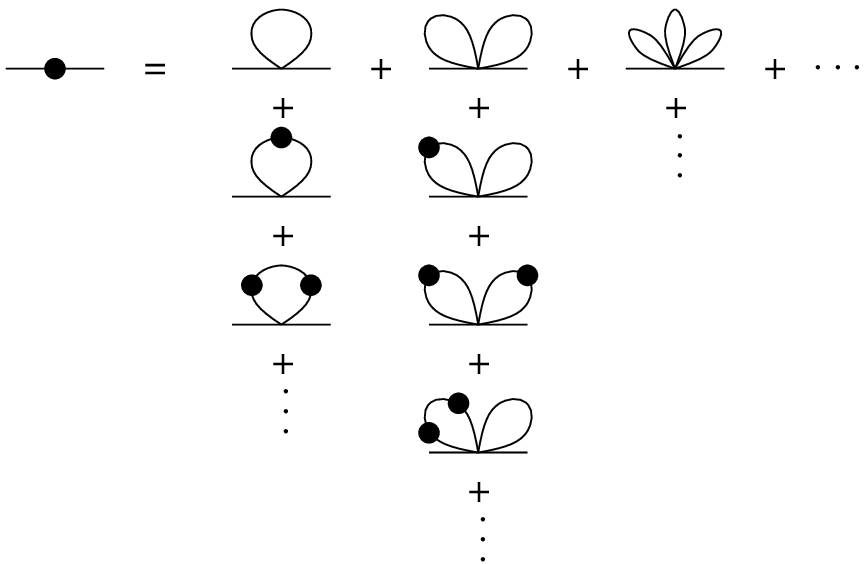,width=10truecm}
\label{recursive}
\end{equation}
Now, the fact that the vertices involved in the above contain only
$F_{x,\mu\nu}^{(1)}$ implies that the longitudinal parts of all
propagators will always cancel. As we will see, this fact will lead to
the result that the effect of dressing is the same in all covariant
gauges. We will thus denote by a thick (thin) solid line the
transverse dressed (bare) propagator.

From Eq.~(\ref{recursive}) there follows:
\begin{equation}
\psfig{figure=propirr1.eps,width=1truecm} = w(g_0) \cdot 
\psfig{figure=propbare.eps,width=1truecm}
\end{equation}
Indeed, the dressed propagator will become a multiple of the bare
transverse one, where the factor $w(g_0)$ will depend on $g_0$ and
$N$, but not on the momentum. Let us now turn the diagrammatic
relations (\ref{propdress}, \ref{recursive}) into an algebraic
equation for $w(g_0)$; from Eq.~(\ref{propdress}) we have:
\begin{equation}
\psfig{figure=propdress.eps,width=1truecm} = 
\psfig{figure=propbare.eps,width=1truecm} \cdot 
\left( 1 + w(g_0) + w(g_0)^2 + \cdots \right) = 
\psfig{figure=propbare.eps,width=1truecm} \cdot 
{1\over 1-w(g_0)}
\label{propdr}
\end{equation}
and from Eq.~(\ref{recursive}) we find:
\begin{equation}
\psfig{figure=propbare.eps,width=1truecm} \cdot w(g_0) =
\psfig{figure=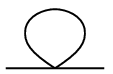,width=1truecm} \cdot {1\over 1-w(g_0)} +
\psfig{figure=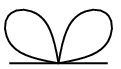,width=1truecm} \cdot {1\over [1-w(g_0)]^2} +
\psfig{figure=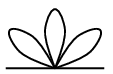,width=1truecm} \cdot {1\over [1-w(g_0)]^3} +
\cdots
\label{recursivew}
\end{equation}
It is crucial to verify at this stage that all diagrams contained
above appear with the same combinatorial factors as in the ordinary
perturbative expansion; this is indeed the case.

To proceed, we must evaluate the generic tadpole appearing in
Eq.~(\ref{recursivew}); this comes from an $n$-point vertex of the
action, in which $n-2$ lines have been pairwise contracted. Before
contraction, the vertex reads:
\begin{eqnarray}
-S &\to& {1\over n!\, g_0^2}\, \sum_{x,\mu\nu} (i g_0)^n\, {\rm
tr}\left\{\left(F_{x,\mu\nu}^{(1)}\right)^n\right\} \nonumber \\
&=& {(i g_0)^n \over n! \, g_0^2} \,\sum_{x,\mu\nu} \int dq_1 \ldots
dq_n \, \left({\hat q}_{1\mu} A_\nu^{a_1}(q_1) - {\hat q}_{1\nu}
A_\mu^{a_1}(q_1)\right) \ldots \left({\hat q}_{n\mu} A_\nu^{a_n}(q_n)
- {\hat q}_{n\nu} A_\mu^{a_n}(q_n)\right) \nonumber \\
&& \qquad\qquad\qquad \cdot e^{i(q_1 + \cdots + q_n)x} \, {\rm tr} \{T^{a_1}
T^{a_2} \ldots T^{a_n} \}
\label{vertexn}
\end{eqnarray}
(${\hat q}_\mu = 2 \sin(q_\mu/2)\,$). At contraction there will be
$(n-2)/2$ loop integrations giving:
\begin{equation}
{1\over (2\pi)^4} \int d^4q\, {2 {\hat q}_\mu^2 \over {\hat q}^2} =
{1\over 2}
\label{loopintegral}
\end{equation}

For the contraction of the $SU(N)$ generators we first define and
evaluate $F(n;N)$, which is the sum over all complete pairwise
contractions of ${\rm tr}\{T^{a_1} T^{a_2}\ldots T^{a_n}\}$:
\begin{equation}
F(n;N) = {1\over 2^{n/2} (n/2)!} \sum_{P\in S_n} \delta_{a_1 a_2}\,
\delta_{a_3 a_4} \ldots \delta_{a_{n-1} a_n}\, {\rm tr} 
\left\{T^{P(a_1)}\, T^{P(a_2)} \ldots T^{P(a_n)} \right\}
\end{equation}
($F(2n+1;N)\equiv 0$; $S_n$ is the permutation group of $n$ objects).
In Appendix~\ref{appG} we calculate the
generating function of $F(n;N)$:
\begin{equation}
G(z;N) \equiv \sum_{n=0}^\infty {z^n\over n!} F(n;N), \qquad {\rm
whence:} \qquad F(n;N) = {d^n\over dz^n} G(z;N)\big|_{z=0}
\label{GzN}
\end{equation}
We find:
\begin{equation}
G(z;N) = e^{z^2 (N-1)/(4N)} \, L^1_{N-1}(-z^2/2)
\label{Gresult}
\end{equation}
($L^\alpha_\beta\,$ are Laguerre polynomials). In the present case,
two legs are left external, so that the color contraction gives:
\begin{equation}
{n \, F(n;N)\over 2( N^2 {-}1)}
\label{contraction2}
\end{equation}
Substituting~(\ref{loopintegral}) and~(\ref{contraction2}) in
Eq.~(\ref{vertexn}), we obtain for the tadpole:
\begin{eqnarray}
{(i g_0)^n \over n! \, g_0^2} &&\sum_{x,\mu\nu} \int dq_1
dq_2 \, ({\hat q}_{1\mu} A_\nu^{a}(q_1) {-} {\hat q}_{1\nu}
A_\mu^{a}(q_1))({\hat q}_{2\mu} A_\nu^{a}(q_2)
{-} {\hat q}_{2\nu} A_\mu^{a}(q_2)) \, e^{i(q_1 + q_2)x}
\, {n\, F(n;N)\over 2 (N^2 {-}1)}\,
({1\over 2})^{n{-}2\over 2}  \nonumber \\
&&= \psfig{figure=propbare.eps,width=1truecm} \cdot 
{2(i g_0)^n \over (n-1)! \, g_0^2}  \cdot {1\over N^2 {-}1} \cdot F(n;N)
\cdot ({1\over 2})^{n{-}2\over 2}
\label{tadpole}
\end{eqnarray}
We can now sum up all terms in Eq.~(\ref{recursivew}); we obtain:
\begin{eqnarray}
w(g_0) &=& \sum_{n= 4,6,8,\ldots} {1\over [1 -w(g_0)]^{(n-2)/2}} \cdot
{2(i g_0)^n \over (n-1)! \, g_0^2}  \cdot {1\over N^2 {-}1} \cdot F(n;N)
\cdot ({1\over 2})^{(n-2)/2} \label{wg0} \\
&=& \left\{\sum_{n=0}^\infty {1\over [2 -2w(g_0)]^{n/2}} \cdot
{(i g_0)^n \over n!} \cdot F(n+1;N) \right\}
\cdot {2 (i g_0)\over g_0^2\, (N^2 {-}1)} 
\cdot [2-2w(g_0)]^{1/2} + 1 \nonumber 
\end{eqnarray}
Comparing with the definition of $G(z;N)$, Eq.~(\ref{GzN}), we see
that the expression in curly brackets above is simply $G'(z;N)$, the
derivative of $G(z;N)$. Eq.~(\ref{wg0}) now reads:
\begin{equation}
zG'(z;N)\big|_{\displaystyle z={(i g_0)/ (2 -2 w(g_0))^{1/2}}} = - {g_0^2 \,
(N^2{-}1)\over 4}
\label{zG}
\end{equation}
From our result for $G(z;N)$, Eq.~(\ref{Gresult}), we see that:
\begin{equation}
zG'(z;N) = e^{z^2 (N{-}1)/(4N)} \,\left[ -{N{-}1\over N} \, L^1_{N{-}1}
(-{z^2\over 2}) -2 L^2_{N{-}2} (-{z^2\over 2}) \right] \,(-{z^2\over 2})
\end{equation}
This allows us to make explicit Eq.~(\ref{zG}):
\begin{eqnarray}
&&u \, e^{-u (N{-}1)/(2N)} \,\left[ {N{-}1\over N} \, L^1_{N{-}1}(u) +2
L^2_{N{-}2}(u) \right] = {g_0^2\,(N^2{-}1)\over 4}, \label{zGeq}\\
&&u(g_0) \equiv {g_0^2 \over 4 (1{-}w(g_0))}\nonumber
\end{eqnarray}

Given $g_0$, $N$, this equation can be solved numerically for $u(g_0)$
and, subsequently, for $w(g_0)$.
The region in $g_0$ for which a
solution exists contains the whole range of physical interest.
Indeed one finds a solution in the region
$0\leq g_0^2 \leq 16/3e^{1/2}\simeq 3.23$
for $N=2$,
and $0\leq g_0^2 \leq 1.558$ for $N=3$.
In Figures 1 and 2 we plot the left hand side of Eq.~(\ref{zGeq}) versus $u$,
for $SU(2)$ and $SU(3)$ respectively.

\subsection{Vertices from the action}
\label{sec2.2}

The 3-point vertex of the action can be dressed to all orders in a
manner similar to Eq.~(\ref{recursive}). We have:
\begin{equation}
\psfig{figure=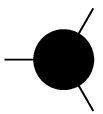,width=1truecm} \ \  
{= \atop \phantom{0}} \ \ 
\psfig{figure=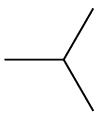,width=1truecm}\ \  {+ \atop \phantom{0}} \ \ 
\psfig{figure=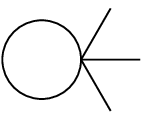,width=1truecm}\ \  {{\displaystyle \cdot 
\  {1\over
1-w(g_0)}  + }\atop \phantom{0}}
\ \ 
\psfig{figure=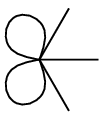,width=1truecm} {{\displaystyle \cdot \ {1\over
[1-w(g_0)]^2} + \cdots} \atop \phantom{0}}
\label{threedressed}
\end{equation}
The calculation is described in Appendix~\ref{app3pt}. The result
turns out to be very simple:
\begin{equation}
\psfig{figure=threedressed.eps,width=1truecm} \ \ {= \atop \phantom{0}}  \ \ 
\psfig{figure=threebare.eps,width=1truecm}\  {{\displaystyle \cdot \ 
(1-w(g_0)) }\atop 
\phantom{0}} 
\label{THREEPOINT}
\end{equation}
where $w(g_0)$ is the quantity calculated previously.

Vertices with more lines can be treated similarly; however, the dressed vertex
in these cases is not merely a multiple of the bare one, which tends
to complicate matters somewhat. Since we will not need such vertices
for the numerical results of Section~\ref{sec3}, we only present some
relevant formulae in Appendic~\ref{app4pt}.

\subsection{Other operators}
\label{sec2.3}

Various lattice operators can be dressed by the same procedure.
Let us take as an example a typical operator
involving gluons:
\begin{equation}
{\cal O} = \sum {\rm tr}\{U_1 U_2\ldots U_n\} \equiv \sum {\rm tr} 
\left\{e^{i g_0 Q}\right\}
\end{equation}
where the sum runs over the Lorentz indices involved. Using the first
order BCH expansion for $Q$, we can write once again for the 2-point
tadpole built out of an $n$-point vertex:
\begin{equation}
\psfig{figure=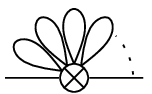,width=1.5truecm}
=
\psfig{figure=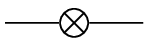,width=1.5truecm}
\cdot {4 (i g_0)^{n-2} \over n!} \cdot {n\, F(n;N)\over 2 (N^2 {-}1)}
\cdot \alpha^{(n-2)/2}
\label{optadpole}
\end{equation}
Here, $\alpha$ is the value of the one-loop momentum integration
coming from the contraction of $Q$ with itself; it is a pure number
which depends on the operator under consideration. For example:
\begin{eqnarray}
{\cal O} &=& \,S,\qquad\qquad\qquad\qquad\qquad\qquad \alpha = {1\over 2},\
{\rm as\ before}\\ 
{\cal O} &=& \sum_{\mu,\nu,\rho,\sigma}
\varepsilon^{\mu\nu\rho\sigma} {\rm tr} \{U_{x,\mu\nu}
U_{x,\rho\sigma} \}, \qquad\, \alpha = 1 \label{topcharge}\\
{\cal O} &=& \sum_{\mu,\nu,\rho} {\rm tr} \{U_{x,\mu\nu}
[U_{x,\nu\rho}, U_{x,\rho\mu}]\}, \qquad \alpha = {3\over 2}  - 
{3\over 2(2\pi)^4} \int d^4q\, { {\hat q}_\mu^2 {\hat q}_\nu^2 \over
{\hat q}^2} \simeq 0.85332
\end{eqnarray}
and so on. The complete resummation of cactus diagrams then leads to:
\begin{eqnarray}
\psfig{figure=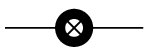,width=1.5truecm}
&=&
\psfig{figure=opbare.eps,width=1.5truecm} +
\psfig{figure=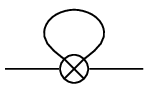,width=1.5truecm} \cdot {1\over 1-w(g_0)} +
\psfig{figure=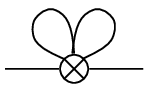,width=1.5truecm} \cdot {1\over [1-w(g_0)]^2} +
\cdots  \label{opdressed} \\
&=& \psfig{figure=opbare.eps,width=1.5truecm} \cdot e^{-x(N-1)/(2N)}
\cdot\left[ 
{N{-}1\over N} \, L^1_{N{-}1}(x) +2 L^2_{N{-}2}(x) \right]_{x = (g_0^2
\alpha)/(2-2w(g_0))} \cdot {1\over N^2{-}1} \nonumber
\end{eqnarray}
It turns out that 3-point bare and dressed vertices are related by the
same proportionality factor as the 2-point vertices,
Eq.~(\ref{opdressed}).

For the topological charge density operator of Eq.~(\ref{topcharge})
an alternative resummation is possible by using the BCH expansion as
follows:
\begin{equation}
\sum_{\mu,\nu,\rho,\sigma}
\varepsilon^{\mu\nu\rho\sigma} {\rm tr} \{U_{x,\mu\nu}
U_{x,\rho\sigma} \} =
\sum_{\mu,\nu,\rho,\sigma}
\varepsilon^{\mu\nu\rho\sigma} {\rm tr} \{\exp (i g_0 F_{x,\mu\nu})
\, \exp (i g_0 F_{x,\rho\sigma}) \}
\end{equation}
Keeping the first order terms in $F_{x,\mu\nu}$ and $F_{x,\rho\sigma}$
the complete resummation leads to the simple result:
\begin{equation}
\psfig{figure=opdressed.eps,width=1.5truecm}
= \psfig{figure=opbare.eps,width=1.5truecm}
\cdot [ 1-w(g_0)]^2
\label{topdressed}
\end{equation}
The square in the above expression can be traced to the fact that the
operator is composed of two mutually othogonal plaquettes.

Cactus resummation can be also used to estimate the perturbative
vacuum expectation value of an operator:
\begin{equation}
\psfig{figure=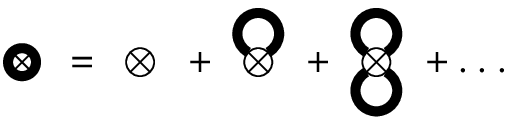,width=8truecm}
\label{vev}
\end{equation}
This can be shown to equal:
\begin{equation}
\psfig{figure=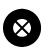,width=0.7truecm} =
G\left({i g_0 \alpha^{1/2} \over [1-w(g_0)]^{1/2}};\, N\right)
\label{vevdressed}
\end{equation}

\subsection{Other representations}
\label{sec2.4}

The calculation performed above can be generalized to encompass
several other cases, e.g. operators involving higher representations
for gluons. To illustrate this, we consider the class of variant
actions proposed some time ago~\cite{B-C}:
\begin{equation}
S = {\beta\over 2} \sum_{x,\mu\nu} \left( 1 - {1\over N}\, {\rm tr}
U_{x,\mu\nu} \right) + {\beta_A\over 2}  \sum_{x,\mu\nu} \left( 1 -
{1\over N^2{-}1} \, {\rm tr}_A U_{x,\mu\nu} \right)
\end{equation}
Here $\beta$ and $\beta_A$ are adjustable parameters and ${\rm tr}_A
U_{x,\mu\nu}$ denotes the trace of a product of links in the adjoint
representation, around a plaquette.

The calculation proceeds as before; the one new ingredient we need is
$F_{\rm Adj}(n;N)$, the sum over all complete pairwise contractions of
${\rm tr}\{{\cal T}^{a_1}{\cal T}^{a_2}\ldots{\cal T}^{a_n}\}$\ \
(${\cal T}^{a}$: generators in the adjoint representation).
In Appendix~\ref{appadjoint} we compute $G_{\rm Adj}(z;N)$, the
generating function for $F_{\rm Adj}\,$. In terms of $G_{\rm Adj}\,$,
the equation for the factor $w_{\rm var}(g_0)$ which dresses the propagator now
becomes: 
\begin{equation}
{\beta\over 2N} \, z G'(z;N) + {\beta_A\over 2(N^2{-}1)} \, z G'_{\rm
Adj}(z;N) \biggr|_{\displaystyle z = 
{(ig_0)\over [2-2w_{\rm var}(g_0)]^{1/2}}} = 
- {N^2{-}1\over 4}
\label{wg0variant}
\end{equation}
where:
\begin{equation}
g_0^2 = \left[{\beta\over 2N} + {\beta_A \, N\over N^2{-}1}\right]^{-1}
\end{equation}
It is straightforward to solve Eq.~(\ref{wg0variant}) numerically for
$w_{\rm var}(g_0)$.

We conclude this Section by noting that the extension to vertices with
fermions is immediate. First of all, vertices coming from the Wilson
fermionic action stay unchanged, since their definition contains no
plaquettes on which to apply the linear BCH formula. We will see how
this affects corresponding renormalization functions in the next
Section. For more complicated fermionic vertices, such as those of the
clover action, cactus resummation proceeds in precisely the same
manner as Eqs.~(\ref{optadpole}, \ref{opdressed}).

\section{Some applications}
\label{sec3}

In this section we apply the resummation of the cactus diagrams
derived for the Wilson action to the calculation of
the renormalization of some lattice operators.
Approximate expressions for these renormalizations on the lattice are
obtained by dressing the corresponding one-loop
results. We will consider here operators whose anomalous dimensions are zero.
A consistent, as well as physically motivated, means of implementing the 
cactus dressing is to apply it to the one-loop 
difference between lattice and 
continuum contributions that determine the renormalization, and not only to
the lattice part. Cases with nonzero anomalous dimension can be dealt with in 
an analogous manner, by setting the scale $\mu = 1/a$ and dressing the finite 
renormalization coefficients as before.

As a first example
we consider the calculation of the lattice renormalization
$Z(g_0^2)$ of the topological charge density operator
\begin{equation}
Q(x) =
-{1\over 2^4\times 32\pi^2}
 \sum_{\mu,\nu,\rho,\sigma=\pm 1}^{\pm 4}
\varepsilon^{\mu\nu\rho\sigma} {\rm tr} \{U_{x,\mu\nu}
U_{x,\rho\sigma} \}.
\label{topchargeb}
\end{equation}
$Z(g_0^2)$ is a finite function of $g_0^2$;
it approaches one in the limit $g_0\to 0$, and is much
smaller than one in the region $g_0\simeq 1$, where 
Monte Carlo simulations
using the Wilson action are actually performed.
A non-perturbative numerical
calculation using the heating method~\cite{D-V} 
has produced the estimate~\cite{Spin,Vi} 
\begin{equation}
Z(g_0^2=1)=0.19(1)\qquad {\rm for} \quad SU(3).
\label{hm}
\end{equation}
In this case few terms of the perturbation theory in $g_0^2$
can hardly provide  an acceptable
estimate of $Z(g_0^2)$ for $g_0^2\simeq 1$
without some kind of resummation.

In perturbation theory $Z(g_0^2)$ has been computed
to $O(g_0^2)$~\cite{C-D-P} 
with the result
\begin{eqnarray}
Z(g_0^2)&=& 1+z_1g_0^2+O(g_0^4), \label{stopz}\\
z_1&=& N\left( {1\over 4N^2} - {1\over 8} - {1\over 2\pi^2}-0.15493
\right).\nonumber 
\end{eqnarray}
Numerically  $z_1\simeq -0.908$ for
$SU(3)$, and $z_1\simeq -0.536$ for
$SU(2)$.
So perturbation theory to $O(g_0^2)$ would give $Z(g_0^2)\simeq 0.092$
for $SU(3)$, which is very far from  its actual value, Eq.~(\ref{hm}).
In order to obtain a better approximation, we perform
a cactus dressing of the one-loop calculation.
The tree order is dressed by mere use of Equation (\ref{topdressed}).
One can now dress the one-loop contributions (for the
details of the standard
perturbative calculation see Refs.~\cite{C-D-P,A-V}).
Using 
Eqs.~(\ref{propdr}), (\ref{THREEPOINT}) and (\ref{topdressed}),
and a simple combinatorial counting applied to the diagrams contributing
to $Z(g_0^2)$, one arrives at the following expression
\begin{equation}
Z_(g_0^2)\approx \left[ 1 - w(g_0)\right]^2 +  
\left[ 1 - w(g_0)\right] \left( z_1 + {2N^2-3\over 12N}\right)g_0^2.
\label{cactusapprox}
\end{equation}
The quantity $(2N^2-3)/12N$ must be added to $z_1$ to avoid double counting, 
since such a contribution is already incorporated in the 
dressed tree-order approximation.
Solving Eq.~(\ref{zGeq}) for $N=3$ and $g_0=1$ one finds
\begin{equation}
1-w(g_0=1)=0.749775.
\end{equation}
Thus from Eq.~(\ref{cactusapprox}) one obtains
$Z(g_0^2=1)\simeq 0.193$, which compares very well with
the numerical result (\ref{hm}).
Further confirmation of the validity of the approximation ~(\ref{cactusapprox})
comes from a comparison with available data for $SU(2)$
in the range $2.45\leq \beta \leq 3.0$ ($\beta=4/g_0^2$)
obtained by the heating method~\cite{A-C-D-G-V}, as
shown in Table~\ref{datasu2}.
The agreement is remarkable.

We wish to point out that other improvement recipes, such as those proposed
in Ref.~\cite{lepage}, consisting in a redefinition of
the bare-coupling, do not help in this case.
For example  one recipe entails use of 
\begin{equation}
\tilde{g}^2 = 
{g_0^2\over \case{1}{3} \langle {\rm Tr} \;U^\Box \rangle },
\end{equation}
(where $U^\Box$ is the plaquette) as the expansion parameter
for $N=3$. In many cases this recipe represents an improvement.
However,  substituting 
the value of $\widetilde{g}^2$ corresponding
to $g_0^2=1$, i.e. $\widetilde{g}^2\simeq 1.68$, in Eq.~(\ref{stopz}),
one would obtain $Z(g_0^2=1)\simeq -0.54$,
that is much worse than the plain one-loop approximation.
Similarly, a change of coupling constant and momentum scale, in the manner of 
Lepage and Mackenzie~\cite{lepage}, also leads to a wider discrepancy in this 
case: indeed, the corresponding value of $\alpha(q^*)$ 
(defined in \cite{lepage}),
turns out to be too large.

\medskip
One can also apply cactus resummation to the lattice renormalization
of fermionic operators. Let us consider the local non-singlet vector and axial
currents $V_\mu^a=\bar{\psi} \lambda^a\gamma_\mu \psi$ and
$A_\mu^a=\bar{\psi} \lambda^a\gamma_\mu \gamma_5\psi$.
The lattice renormalizations of these operators, $Z_V(g_0^2)$ and
$Z_A(g_0^2)$ respectively,  
are again finite functions of $g_0$.
In perturbation theory  and for $SU(3)$
one has~\cite{M-Z} 
\begin{equation}
Z_{V,A} = 1 + z_{V,A} g_0^2 + O(g_0^4),
\end{equation}
where $z_V\simeq -0.17$ and $z_A\simeq -0.13$.
Thus, at $g_0^2=1$ one-loop perturbation theory
gives  $Z_V(g_0^2=1)\simeq 0.83$ 
and $Z_A(g_0^2=1)\simeq 0.87$. 
For these fermionic operators,
one may use cactus resummation to
dress the gluon propagators appearing in the diagrams contributing
to one-loop order, according to Eq.~(\ref{propdr}).
This procedure leads to
\begin{equation}
Z_{V,A} \approx 1 + z_{V,A}\, {g_0^2\over 1-w(g_0)}. 
\end{equation}
At $g_0^2=1$, this gives:
$Z_V\simeq 0.77$ and 
$Z_A\simeq 0.83$. 
One may compare these numbers with those obtained
in nonperturbative
calculations based on the Ward identities~\cite{romani}.
The only limitation of this method is due to scaling corrections,
which turn out
to be rather large at $g_0^2\simeq 1$ in the case of the Wilson 
lattice formulation. 
Depending on the matrix element one looks at, at $g_0^2=1$ one finds
values ranging from 0.57 to 0.79 for $Z_V$ and
from 0.72 to 0.85 for $Z_A$~\cite{M-M,Sachrajda,Gock}
(see also Ref.~\cite{C-L-V} for a review of these results).
Other methods of improvement (see, e.g.,~\cite{lepage}, also 
~\cite{C-L-V} for a partial review), using various boosting procedures, 
result in
numbers ranging from 0.63 to 0.71 for $Z_V$ and
from 0.72 to 0.77 for $Z_A$.
Hence, a conclusive comparison is not possible in these cases.
A better numerical situation occurs when one considers
the clover action~\cite{clover}, for which 
scaling corrections are largely reduced in the
region where Monte Carlo simulations are performed and
precise estimates can be obtained using the Ward identities
(see Ref.~\cite{C-L-V} and references therein).
An application  of our cactus
resummation to the clover action 
would require the dressing of the new
fermion-gluon three-point vertex. This point is under investigation. 

In conclusion, the above examples show that the resummation of cactus 
diagrams leads to a general improvement in the evaluation of
lattice renormalizations based on perturbation theory. A combination of this 
method with improved actions is expected to give a reliable evaluation of 
renormalization functions, which can complement corresponding nonperturbative 
estimates. We hope to return to this issue in a future publication.

\acknowledgements

It is a pleasure to thank Paolo Rossi for very useful
discussions. H. P. would like to acknowledge the warm
hospitality extended to him by the Theory Group in Pisa during various
stages of this work.

\appendix

\section{Calculation of $G(\ZETA;N)$}
\label{appG}

We wish to calculate $F(n;N)$, the sum over all complete pairwise
contractions of ${\rm tr}\{T^{a_1} T^{a_2}\ldots T^{a_n}\}$. For even
$n$, $F(n;N)$ is defined by:
\begin{equation}
F(n;N) = {1\over 2^{n/2} (n/2)!} \sum_{P\in S_n} \delta_{a_1 a_2}\,
\delta_{a_3 a_4} \ldots \delta_{a_{n-1} a_n}\, {\rm tr} 
\left\{T^{P(a_1)}\, T^{P(a_2)} \ldots T^{P(a_n)} \right\}
\end{equation}
($F(2n+1;N)\equiv 0$). $S_n$ is the permutation group of $n$ objects,
and $T^a$ are an orthonormal basis for $su(N)$ in the fundamental
representation, ${\rm tr}\{T^a T^b\} = {1\over 2} \delta^{ab}$.

We define:
\begin{equation}
M\equiv \theta^a T^a, \quad \theta^2 = \theta^a \theta^a \quad (a =
1,\ldots , N^2{-}1), \quad \theta^a \in {\real} 
\end{equation}
Then $F(n;N)$ can be written as:
\begin{equation}
F(n;N) = {1\over {\cal N}} \int \prod_a d\theta^a \, e^{-{1\over 2}
\theta^2} \, {\rm tr}\{M^n\}
= {1\over {\cal N}} \int [dM] \, e^{-M^2} \, {\rm tr}\{M^n\}, \qquad
{\cal N} = \int [dM] \, e^{-M^2}
\label{FnN}
\end{equation}
(The normalization ${\cal N}$ is redefined below as necessary, to
ensure that $F(0;N) = N$ remains valid). 
By definition, $[dM] = \prod_a d\theta^a$ is the integration measure
over traceless Hermitian matrices. When the integrand is invariant
under similarity transformations, as is our case, ``angular''
integrations can be performed, leaving behind an integral over the $N$
eigenvalues $\lambda_i$~\cite{mehta}:
\begin{equation}
F(n;N) = {1\over {\cal N}} \int (\prod_i d\lambda_i)\, \Big[\prod_{i<j}
(\lambda_i - \lambda_j)^2\Big] \, \delta
(\sum_i \lambda_i) \, e^{-\sum_i \lambda_i^2} 
\, \left(\sum_i \lambda_i^n \right)
\label{Flambda}
\end{equation}

At this stage, it is convenient to introduce the generating function
for $F(n;N)$:
\begin{equation}
G(z;N) \equiv \sum_{n=0}^\infty {z^n\over n!} F(n;N), \qquad 
\qquad F(n;N) = {d^n\over dz^n} G(z;N)\big|_{z=0}
\end{equation}
By Eq.~(\ref{Flambda}) we have:
\begin{eqnarray}
G(z;N) &=& \sum_i \int {\prod_j d\lambda_j \over {\cal
N}} \, \Big[\prod_{k<l}
(\lambda_k - \lambda_l)^2\Big] \, \delta (\sum_m \lambda_m) \,
e^{-\sum_n \lambda_n^2 + z \lambda_i} \nonumber \\ 
&=& N \int {\prod_j d\lambda_j \over {\cal N}} \, \left[\prod_{k<l}
(\lambda_k - \lambda_l)^2\right] \, \delta (\sum_m \lambda_m) \,
e^{-\sum_n \lambda_n^2 + z \lambda_1} 
\label{Glambda}
\end{eqnarray}
To simplify the exponents we shift the $\lambda$'s so as to keep their
sum equal to zero:
\begin{eqnarray}
\lambda'_1&=& \lambda_1 + {z\over 2N} - {z\over 2} \nonumber \\
\lambda'_i&=& \lambda_i + {z\over 2N} \qquad\qquad (i\ne 1)
\end{eqnarray}
\begin{equation}
G(z;N) = 
N \int {\prod_i d\lambda_i \over {\cal N}} \, \Big[\prod_{1\ne k<l}
(\lambda_k - \lambda_l)^2\Big]  \Big[\prod_{k\ne 1}
(\lambda_1 + {z\over 2} - \lambda_k)^2\Big] \, \delta (\sum_i \lambda_i) \,
e^{-\sum_i \lambda_i^2 + z^2 (N-1)/(4N)} 
\end{equation}
The $\delta$-function can now be easily eliminated, using the
exponential representation $\int d\alpha \exp (i \alpha \sum_i
\lambda_i)$:
\begin{eqnarray}
G(z;N) &=&
N \, e^{z^2 (N-1)/(4N)} \int {d\alpha \, \prod_i d\lambda_i \over {\cal N}} \,
\Big[\prod_{1\ne k<l} 
(\lambda_k {-} \lambda_l)^2\Big] \Big[\prod_{k\ne 1}
(\lambda_1 {+} {z\over 2} {-} \lambda_k)^2\Big] \, 
e^{-\sum_i (\lambda_i - i\alpha/2)^2{-}\alpha^2 N/2} \nonumber
\\
&=& N \, e^{z^2 (N-1)/(4N)} \int {\prod_i d\lambda_i \over {\cal N}} \,
\Big[\prod_{1\ne k<l} 
(\lambda_k {-} \lambda_l)^2\Big] \, \Big[\prod_{k\ne 1}
(\lambda_1 {+} {z\over 2} {-} \lambda_k)^2\Big] \, 
e^{-\sum_i \lambda_i^2} 
\end{eqnarray}
Let us isolate the integral over $\lambda_i\ (i\ne 1)$:
\begin{equation}
G(z;N) = 
N \, e^{z^2 (N-1)/(4N)} \int d\lambda_1 \, e^{-\lambda_1^2} \,
e^{(\lambda_1+ z/2)^2}
 \Big[\int {\prod_{i\ne 1} d\lambda'_i \over {\cal N}} \,
\big(\prod_{k<l} 
(\lambda'_k {-} \lambda'_l)^2\big)  
e^{-\sum_i {\lambda'_i}^2} \Big]_{\lambda'_1 = \lambda_1
+ {z\over 2}}
\end{equation}
The integral in square brackets, involving the Vandermonde determinant
$\prod_{k<l} (\lambda'_k - \lambda'_l)$, equals~\cite{mehta}:
\begin{equation}
{1\over N} \, \sum_{j=0}^{N{-}1} \phi_j^2 \left(\lambda_1 + {z\over
2} \right), \qquad
\phi_j(x) \equiv (2^j\, j!\, \sqrt{\pi})^{-1/2} \, e^{x^2/2}
\left(-{d\over dx}\right)^j \, e^{-x^2}
\label{intlambda}
\end{equation}
We thus obtain:
\begin{eqnarray}
G(z;N)&=& e^{z^2 (N-1)/(4N)} \,\int d\lambda_1 \, e^{-\lambda_1^2} \,
e^{(\lambda_1+ z/2)^2} \, \sum_{j=0}^{N{-}1} {1\over 2^j\, j!\,
\sqrt{\pi}}\, e^{(\lambda_1+ z/2)^2} \, \left[\left( - {d\over
d\lambda_1} \right)^j \, e^{-(\lambda_1+ z/2)^2} \right]^2 \nonumber
\\
&=& e^{z^2 (N-1)/(4N)} \,\int d\lambda_1 \, e^{-\lambda_1^2} \,
\sum_{j=0}^{N{-}1} {1\over 2^j\, j!\,
\sqrt{\pi}}\, \left[H_j\left(\lambda_1 + {z\over 2}\right)\right]^2 \nonumber
\\
&=& e^{z^2 (N-1)/(4N)} \,\sum_{j=0}^{N{-}1} L_j^0\left(-{z^2 \over 2}
\right) \nonumber \\
&=& e^{z^2 (N-1)/(4N)} L_{N{-}1}^1\left(-{z^2 \over 2} \right)
\label{intlambda1}
\end{eqnarray}
in terms of the Hermite ($H_j$) and Laguerre ($L^\alpha_\beta$) polynomials.

\section{Proof of Eq.~(\ref{threepoint})}
\label{app3pt}

To prove Eq.~(\ref{THREEPOINT}), we must first evaluate the $j$-loop
tadpole diagrams appearing in Eq.~(\ref{threedressed}). Contracted
legs come from $F_{x,\mu\nu}^{(1)}$, while external legs necessarily
originate from $F_{x,\mu\nu}^{(1)}$ and $F_{x,\mu\nu}^{(2)}\,$.
The corresponding vertex comes from
\begin{equation}
-{1\over g_0^2} \sum_{\mu\nu} {\rm tr} \left(1- \exp(i g_0
F_{x,\mu\nu})\right)
\end{equation}
taking $2j+2$ powers from the exponent. We have:
\begin{equation}
\psfig{figure=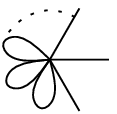,width=1truecm} \ \  
{= \atop \phantom{0}} \ \ 
\psfig{figure=threebare.eps,width=1truecm} \ \  
{{\displaystyle \cdot {2! (i g_0)^{2j} \over (2j+2)!} \cdot \left[
{2j+2 \over N^2 {-}1} 
\, F(2j+2;N) \right] \cdot \left({1\over 2}\right)^j} \atop \phantom{0}}
\label{threetadpole}
\end{equation}
The first factor above is the ratio of Taylor coefficients for the
vertices on the l.h.s. and r.h.s.; the factor in square brackets is
the outcome of color contractions; the factor $(1/2)^j$ is the outcome
of $j$ 1-loop integrations. Combining Eqs.~(\ref{threetadpole})
and~(\ref{threedressed}) we find:
\begin{eqnarray}
\psfig{figure=threedressed.eps,width=1truecm} \ \  
&&{= \atop \phantom{0}} \ \ 
\psfig{figure=threebare.eps,width=1truecm} \ \  
{{\displaystyle \cdot \left\{ \sum_{n=0}^\infty {1\over n!}\, {(i g_0)^n
\over [2-2w(g_0)]^{n/2}} \, {d^n\over dz^n}G'(z;N)|_{z=0} \right\}\cdot
{2[2-2w(g_0)]^{1/2}\over (ig_0)(N^2{-}1)} } \atop \phantom{0}}
\nonumber \\
&&{= \atop \phantom{0}} \ \ 
\psfig{figure=threebare.eps,width=1truecm} \ \  
{{\displaystyle \cdot \, G'({ig_0\over [2-2w(g_0)]^{1/2}};N)\cdot
{2[2-2w(g_0)]^{1/2}\over (ig_0)(N^2{-}1)} } \atop \phantom{0}}
\nonumber \\
&&{= \atop \phantom{0}} \ \ 
\psfig{figure=threebare.eps,width=1truecm} \ \  
{{\displaystyle \cdot \, [1-w(g_0)]} \atop \phantom{0}}
\end{eqnarray}
In the last equality, use was made of Eq.~(\ref{zG}).

\section{Dressing the 4-point vertex}
\label{app4pt}

The bare 4-point vertex contains parts coming from ${\rm
tr}\{F_{x,\mu\nu}^{(1)} F_{x,\mu\nu}^{(1)} F_{x,\mu\nu}^{(1)}
F_{x,\mu\nu}^{(1)} \}$, ${\rm tr}\{F_{x,\mu\nu}^{(2)}
F_{x,\mu\nu}^{(2)} \}$, ${\rm tr}\{F_{x,\mu\nu}^{(1)}
F_{x,\mu\nu}^{(3)} \}$. In general, these are expected to dress
differently, thus yielding a dressed vertex which is not merely
proportional to the bare one. In all other respects, this calculation
is a direct generalization of the 3-point vertex case.

We will not present the final expression for the dressed 4-point
vertex, since we will not be needing it in Section~\ref{sec3}; rather
we evaluate the one new ingredient present in this case: The sum over
all pairwise contractions of ${\rm tr}\left\{(F_{x,\mu\nu}^{(1)})^n\right\}$,
with 4 legs left external. The result can be written as:
\begin{equation}
({1\over 2})^{(n-4)/2} \,\, F_{x,\mu\nu}^{(1)a} F_{x,\mu\nu}^{(1)b}
F_{x,\mu\nu}^{(1)c} F_{x,\mu\nu}^{(1)d} \,\,\, T^{abcd}
\end{equation}
where $T^{abcd}$ is necessarily of the form:
\begin{equation}
T^{abcd} = \alpha\,\cdot\, {\rm tr} \{ T^a (T^b T^c T^d + {\rm permutations})
\} + \beta \, \cdot \, (\delta^{ab} \delta^{cd} + \delta^{ac} \delta^{bd} +
\delta^{ad} \delta^{bc} ) 
\end{equation}
We must compute $\alpha$ and $\beta$ for generic $n$, $N$.

The tensors multiplying $\alpha$ and $\beta$ above are in general
independent, except for the cases $N=2$, $N=3$. One way to see this is
by taking the scalar product of $T^{abcd}$ (a real tensor) with
itself:
\begin{eqnarray}
T^{abcd}T^{abcd} &=& 3(N^2-1)(N^2+1) \left[\alpha^2 \, {N^4 - 6N^2 +18
\over 8N^2 (N^2+1)} + \alpha\beta \, {2N^2-3\over N(N^2+1)} + \beta^2\right]
\nonumber \\
&=& 3(N^2-1)(N^2+1) \left[ (\beta + {2N^2-3\over 2N(N^2+1)} \,
\alpha)^2 + \alpha^2 \, {(N^2-9)(N^2-4)\over 8(N^2+1)^2} \right]
\end{eqnarray}
The above can vanish only for $N=2$ or $N=3$, if $\beta = -{1\over
4}\alpha$.

To compute $\alpha$ and $\beta$ we further contract $T^{abcd}$ with
either ${:}\delta^{ef} \delta^{gh}{:}$ or ${:}{\rm tr}\{T^e T^f T^g T^h\}{:}$,
to arrive at two relations for $\alpha$ and $\beta$:
\begin{eqnarray}
{4! (N^2{-}1)\over 2N} \,[(2N^2 {-}3)\, \alpha + 2N(N^2{+}1)\, \beta] 
&=& n(n{-}2)\, F(n;N)
\label{alphabeta} \\
{4! (N^2{-}1)\over 16N^2}\, [(N^4 {-}6N^2 {+}18)\,\alpha +
4N(2N^2{-}3) \,\beta]  
&=& {n\over 2}\, F(n{+}2;N) -
{n(2N^2{-}4{+}n)\over 4N}\, F(n;N) \nonumber \\
&& \qquad + {n(n{-}1)(N^2{+}n{-}3)\over
8N^2}\, F(n{-}2;N)  
\nonumber
\end{eqnarray}
The solution of these linear equations gives the required expressions for
$\alpha$ and $\beta$. 

For $N=2$, $N=3$, Eqs.~(\ref{alphabeta}) are linearly dependent, as
expected. Since the two tensors in $T^{abcd}$ are proportional to each
other in this case, we can set $\alpha = 0$. Then, from
Eq.~(\ref{alphabeta}) we 
find:
\begin{equation}
\alpha=0, \quad \beta = {n(n-2)\, F(n;N) \over 4!
(N^2-1)(N^2+1)},\qquad (N=2\ {\rm or}\ N=3)
\end{equation}

\section{Results for the adjoint representation}
\label{appadjoint}

We calculate:
\begin{equation}
F_{\rm Adj}(n;N) = {1\over 2^{n/2} (n/2)!} \sum_{P\in S_n} \delta_{a_1 a_2}\,
\delta_{a_3 a_4} \ldots \delta_{a_{n-1} a_n}\, {\rm tr} 
\left\{{\cal T}^{P(a_1)}\, {\cal T}^{P(a_2)} \ldots {\cal T}^{P(a_n)} \right\}
\label{Fadj}
\end{equation}
Here, ${\cal T}^a$ denote $su(N)$ generators in the adjoint
representation. We can relate them to the fundamental representation
using the standard decomposition:
\begin{equation}
(N) \otimes (N^*) \longrightarrow (N^2{-}1) \oplus (1)
\end{equation}
In terms of the generators, this says that there exists a unitary
matrix $U$ such that:
\begin{equation}
U^\dag \,( T^a \otimes \openone + \openone \otimes {T^a}^*) \, U
= {\cal T}^a \oplus (0)
\end{equation}
Using an $N^2 \times (N^2{-}1)$ projector $P$ (the
$(N^2{-}1){\times}(N^2{-}1)$ unit matrix augmented by a row of
zeroes):
\begin{equation}
{\cal T}^a = P^\dag \, U^\dag \,( T^a \otimes \openone + \openone
\otimes (T^a)^*) \, U \, P
\end{equation}
Substituting this in~(\ref{Fadj}) and making use of Eq.~(\ref{FnN}) we
find:
\begin{eqnarray}
F_{\rm Adj}(n;N) &=& {1\over {\cal N}} \int \prod_a d\theta^a \, e^{-{1\over 2}
\theta^2} \, {\rm tr}\left\{ (M\otimes\openone + \openone\otimes
M^*)^n \right\} \nonumber \\
&=& {1\over {\cal N}} \int \prod_a d\theta^a \, e^{-{1\over 2}
\theta^2} \sum_{m=0}^n \left({n\atop m}\right) \, {\rm tr}\{ M^m \} \,
{\rm tr}\{ M^{n-m} \}
\end{eqnarray}

The corresponding generating function is now given by:
\begin{eqnarray}
F_{\rm Adj}(n;N) &=& {d^n\over dz^n} G_{\rm Adj}(z;N)|_{z=0} \nonumber
\\
G_{\rm Adj}(z;N) &=& \sum_{n_1,\, n_2 = 0}^\infty {z^{n_1+n_2}\over
n_1!n_2!} \, {1\over {\cal N}} \int \prod_a d\theta^a \, e^{-{1\over 2}
\theta^2} \, {\rm tr}\{ M^{n_1} \} \, {\rm tr}\{ M^{n_2} \} \nonumber
\\
&=& \sum_{i,j} \int {\prod_m d\lambda_m \over {\cal
N}} \, \Big[\prod_{k<l}
(\lambda_k - \lambda_l)^2\Big] \, \delta (\sum_m \lambda_m) \,
e^{-\sum_n \lambda_n^2 + z (\lambda_i + \lambda_j)}
\end{eqnarray}
A somewhat tedious integration over $\lambda_m$, by analogy with
Eqs.~(\ref{intlambda},\ref{intlambda1}), gives:
\begin{eqnarray}
G_{\rm Adj}(z;N) &=& G(2z;N) + e^{z^2 (N-2)/(2N)} \biggl\{
\left(L^1_{N-1}(-{z^2\over 2})\right)^2 - \sum_{n=0}^{N-1} 
\left(L^0_{n}(-{z^2\over 2})\right)^2 \nonumber \\
&&\qquad -2 \sum_{n=1}^{N-1} \sum_{m=0}^{N-1-n} (-{z^2\over 2})^{n+2m}
\, {1\over m! (n+m)!} \, L_{N-n-m-1}^{n+2m+1}(-{z^2\over 2}) \biggr\}
\end{eqnarray}

From this point on, dressing the variant action propagator proceeds
just as in Eqs.~(\ref{recursivew},\ref{tadpole}), leading to
\begin{eqnarray}
w_{\rm var}(g_0) &=& 
\left\{\sum_{n=0}^\infty {1\over [2 -2w_{\rm var}(g_0)]^{n/2}} \cdot
{(i g_0)^n \over n!} \cdot \left({\beta\over 2N} F(n+1;N) + 
{\beta_A \over 2(N^2-1)} F_{\rm Adj}(n+1;N) \right) \right\} \nonumber
\\
&&\qquad \cdot {2 (i g_0)\over N^2 {-}1} 
\cdot [2-2w_{\rm var}(g_0)]^{1/2} + 1 
\end{eqnarray}

Substituting the definitions of $G(z;N)$ and $G_{\rm Adj}(z;N)$ in the above
produces immediately Eq.~(\ref{wg0variant}).


\break

\centerline{FIGURES}
\medskip
\bigskip
\centerline{\psfig{figure=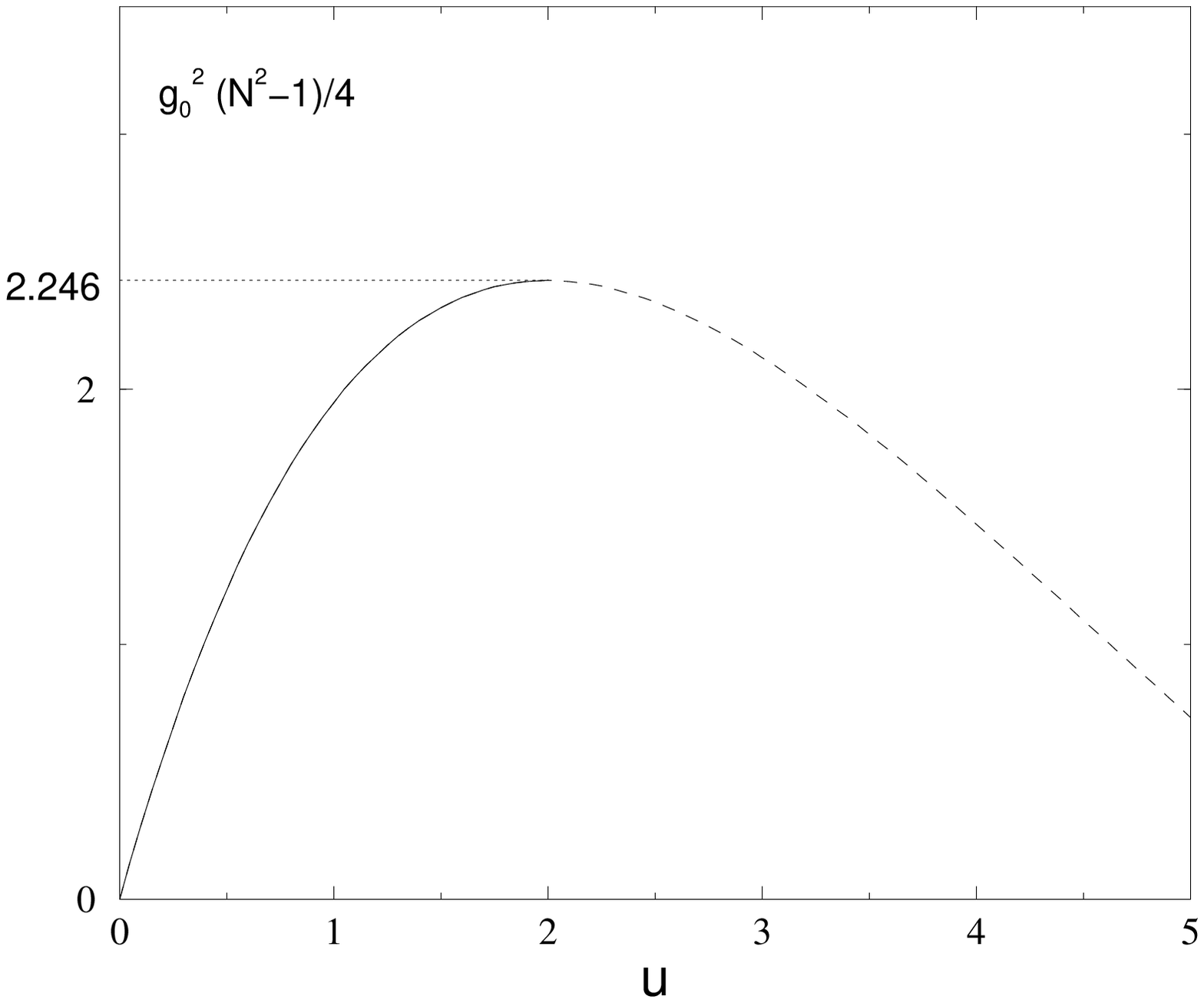,width=10truecm}}
\begin{enumerate}
\item[Figure 1: ] Plot of the l.h.s. of Eq.~(\ref{zGeq}) versus $u$, for 
$SU(2)$. The solid part of the curve identifies the interval of $g_0$ values 
for which a solution exists.
\end{enumerate}
\bigskip\bigskip
\centerline{\psfig{figure=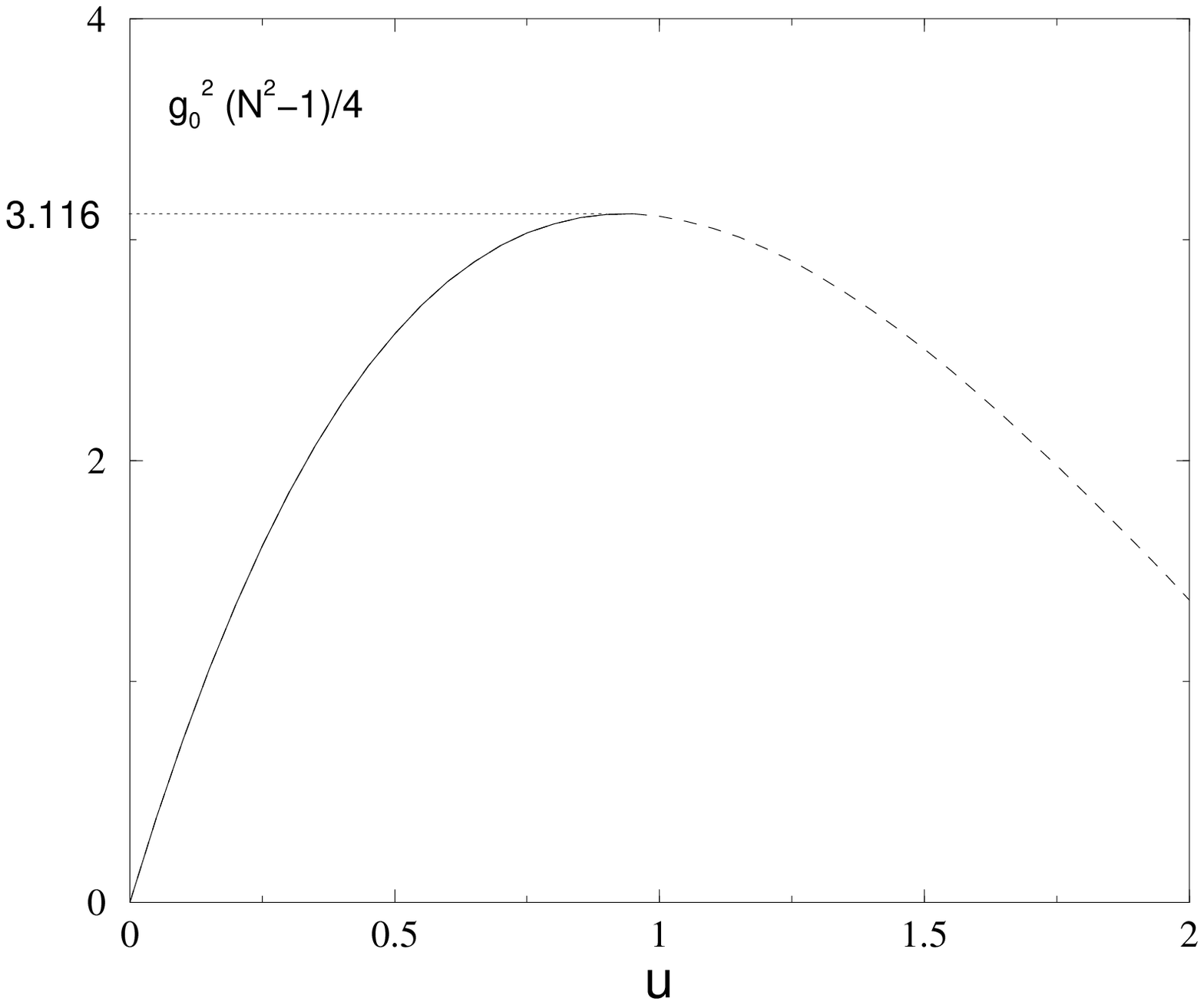,width=10truecm}}
\begin{enumerate}
\item[Figure 2: ] As in Figure 1, for the case of $SU(3)$.

\end{enumerate}


\begin{table}
\caption{
For the $SU(2)$ lattice gauge theory 
we list the estimates of $Z(g_0^2)$ as obtained
by the heating method~\protect\cite{A-C-D-G-V} (h.m.),
by the standard one-loop perturbative expansion (p.t.),
and by cactus dressing, Eq.~(\ref{cactusapprox}),
of the one-loop calculation (d.p.t.).
\label{datasu2}}
\begin{tabular}{r@{}lr@{}lr@{}lr@{}l}
\multicolumn{2}{c}{$\beta\equiv 4/g_0^2$}&
\multicolumn{2}{c}{h.m.}&
\multicolumn{2}{c}{p.t.}&
\multicolumn{2}{c}{d.p.t.}\\
\tableline \hline
2&.45 & 0&.20(2) & 0&.125 & 0&.219 \\
2&.5 & 0&.22(1) & 0&.142 & 0&.233 \\
2&.6 & 0&.25(2) & 0&.175 & 0&.259 \\
2&.8 & 0&.32(2) & 0&.234 & 0&.305 \\
3&.0 & 0&.33(2) & 0&.285 & 0&.347
\end{tabular}
\end{table}

\end{document}